\documentclass[aps,pra,preprint,amsmath,amssymb,showpacs]{revtex4}
\usepackage{epsfig}
\usepackage{times}
\usepackage{mathrsfs}
\usepackage{graphicx}
\usepackage{epsfig}
\usepackage{dcolumn}
\usepackage{color}
\usepackage{bm}
\usepackage{soul}

\begin {document}
\title{Optical cooling of interacting atoms in a tightly confined trap}

\author{Somnath Naskar$^{1,2}$, Subrata Saha$^1$, Partha Goswami$^1$, Arpita pal$^1$ and Bimalendu Deb$^{1,3}$}
\affiliation{$^1$Department of Materials Science, Indian Association
for the Cultivation of Science, India\\
$^2$Department of Physics, Jogesh Chandra Chaudhuri College, Kolkata-700033, India\\
$^3$Raman Center for Atomic, Molecular and Optical Sciences, IACS, Jadavpur, Kolkata 700032, India.}
\begin{abstract}
In a recent paper, we have proposed a novel laser cooling scheme for reducing collisional energy of a pair of atoms by using
photoassociative transitions. In that paper, we considered two atoms in free space, that is we have not considered the effects
of trap on the cooling process. Here in this paper, we qualitatively discuss the possibility of extending this idea for Raman
sideband cooling of a pair of interacting atoms trapped in Lamb-Dicke (LD) regime. Apart from cooling, our method may
be important for manipulating on-site interaction of atoms in an optical lattice.
\end{abstract}

\maketitle

\section{Introduction}
The development of laser cooling and trapping over the last four decades has not only generated renewed interests in
atomic, molecular and optical sciences, but also revitalized the entire gamut of physics. As a result, a number of breakthrough achievements have been possible over the last two decades, most notable being the realization of Bose-Einstein condensation (BEC) in atomic gases \cite{1stbec}, Fermionic superfluidity in ultracold atoms \cite{fermisuperfluidity}, superfluid Mott insulator transition \cite{superfluidmottinsul}, etc.
The idea of laser cooling was first put forward by  H\"{a}nsch and Schawlow \cite{hansch1975}, and also by Wineland and Dehmelt \cite{wineland1975} in 1975.
In 1978, Wineland, Drullinger and Walls cooled \cite{wineland1978} for the first time a cloud of Mg ions held in a penning trap. In the same year Neuhauser, Hohen-Statt, Toschek and 
Dehmelt reported  laser cooling \cite{Neuhauser1978} of trapped Ba$^+$ ions. In 1979, Wineland and Itano \cite{Wineland1979} investigated a cooling process by irradiating atoms with near resonant electromagnetic radiation. An atom moving against the propagation of radiation experiences retardation that results from radiation pressure. In 1968, before the advent of laser cooling, Letokhov \cite{Letokhov1968} proposed trapping of atoms using dipole forces.
   
 Using multiple laser beams in some specific geometric configurations, it is possible to manipulate both internal \cite{kastler} and external degrees-of-freedom (DOF) \cite{kastler,Sc1980,pr1981,RMP1986,phystoday1990,book:lasercooling:metcalf} of slowly moving atoms. One can control 
 the external DOF such as position and momentum of atoms by laser-induced radiative forces which are of two types. One is dissipative force associated with the transfer of linear momentum from the incident light beam to the atom 
in resonant scattering processes.  The other is called dipole force that arises from the interaction between induced dipole moment and the gradient of light field. When two  counter propagating 
laser beams with same frequency and slightly red-detuned from the resonance irradiate moving atoms, an interesting effect 
arises from radiation pressure force. The counter propagating wave comes closer to resonance and exerts a stronger radiation pressure force than what the co-propagating 
wave does. The net force is thus opposite to the atomic velocity. By using three pairs of counter propagating laser waves along three orthogonal directions, one can damp the atomic 
velocity in a very short time, on the order of a few microseconds, achieving what is called an optical molasses\cite{chu1985,Lett1989,chang2014}. 

It was experimentally
 found that the temperature in optical molasses was much lower than expected \cite{Lett1988}, indicating that other laser cooling mechanisms than Doppler cooling exist. The additional damping
 was found to originate from the non-adiabatic response of atoms moving through the light field, the Zeeman structure of the ground
 state and position-dependent optical pumping rates. In fact, there exist sub-Doppler cooling mechanisms resulting from an interplay between spin and external DOF, 
and between dispersive and dissipative effects. One such mechanism is the so-called ‘‘Sisyphus cooling’’ or ‘‘polarization-gradient cooling’’ 
mechanism \cite{cohen-tannoudji:josab,Chu1989} which leads to temperatures much lower than Doppler cooling. The method of Sisyphus cooling of atoms was first proposed by Pritchard 
\cite{prl:1983:pritchard}. It involves an atom having degenerate sub-levels in its 
electronic ground and excited states. It uses counter-propagating polarized lasers to produce spin-dependent and spatially modulated light shifts that eventually 
lead to sub-Doppler cooling.

Laser cooling methods developed so far have utilised electric dipole transitions at a single atom level. This means that laser cooling has so far remained single-atom optical phenomenon. For an interacting atomic sytem there is no optical cooling method for reducing the collisional or interaction energy. To study many-body physics of an interacting atomic gas, particularly to explore quantum degeneracy in an interacting system, it is important to 
devise new method for cooling, trapping and simultaneously manipulating interaction of atomic gases in different 
interaction regimes. We have addressed \cite{pact} a new method by which it is possible to optically cool and trap an interacting pair of atoms.

Recently, we have proposed a novel scheme for photoassociative cooling and trapping (PACT) of a pair of atoms whereby one can reduce collisional energy and trap the 
center-of-mass (CM) motion of the pair \cite{pact}. The scheme is based on photoassociative coupling between an excited molecular bound state and a single-channel 
continuum of states of scattering between two ground state atoms. Photoassociation (PA) is a laser 
induced process where a pair of colliding cold atoms in the electronic ground state interacts with a photon to form a 
quasi-molecule in the excited state. Naturally, both relative and CM motion of the atom-pair become coupled by this process. Thus, it is quite interesting and innovative to 
ask about the possibility of coherent manipulation of this coupled motion and utilize it to cool and trap the pair of atoms. The basic idea of our proposed PACT method \cite{pact} relies on 
Sisyphus effect using photoassociative transitions. In this case the sub-levels correspond to molecular angular momenta associated with multiple scattering 
channels of the atom pair having the same asymptotic threshold and the spin-dependent spatial modulation of light shifts occurs in the CM coordinate space. Thus, here the Sisyphus mechanism 
occurs in the CM coordinate space of the atom-pair. The novelty of this proposal lies in the concept of pair cooling which is fundamentally different from single atom cooling mechanisms. 
Cyclic PA transition between an excited molecular bound state and the continuum of ground-state scattering states is required for cooling. The fulfillment of 
such condition is indeed evidenced in case of narrow-line PA systems from recent experiments \cite{prl:2013:yan,prl:2016:taie}. During the interaction of the atom-pair with a photon, the exchange 
of momentum occurs between the photon and the CM motion since the CM momentum $\left(P_{CM}\right)$ is a good quantum number in such process. Therefore, in a complete cooling cycle, the excess momentum $\Delta P=\hbar\Delta\nu/c$ carried 
out by a photon comes from the CM momentum. This reduces the energy of CM motion by $P_{CM}^2/2M$, $M$ being the total mass of the atom-pair. The excess energy $\Delta E=\hbar\Delta\nu$ carried out 
by the photon, however, comes from the coupled motion of the two atoms as dictated by the principle of energy conservation. Hence, in one cooling cycle, both the CM and 
relative motion is cooled. As the CM momentum remains a good quantum number in a cycle, after multiple cooling cycle the CM motion of the atom-pair can be trapped.

Although unique, the success of the above proposed method is found to depend crucially on the
strength and nodal structure of the free-bound overlap integral at low collisional energy. A class of excited molecular bound states known as purely long-range (PLR) states plays an important role in this scheme. Otherwise, in general, in the presence of an atomic continuum there exists a large no of incoherent decay channels which hinders the PACT mechanism. 

\begin{figure}
 \includegraphics[width=0.8\textwidth]{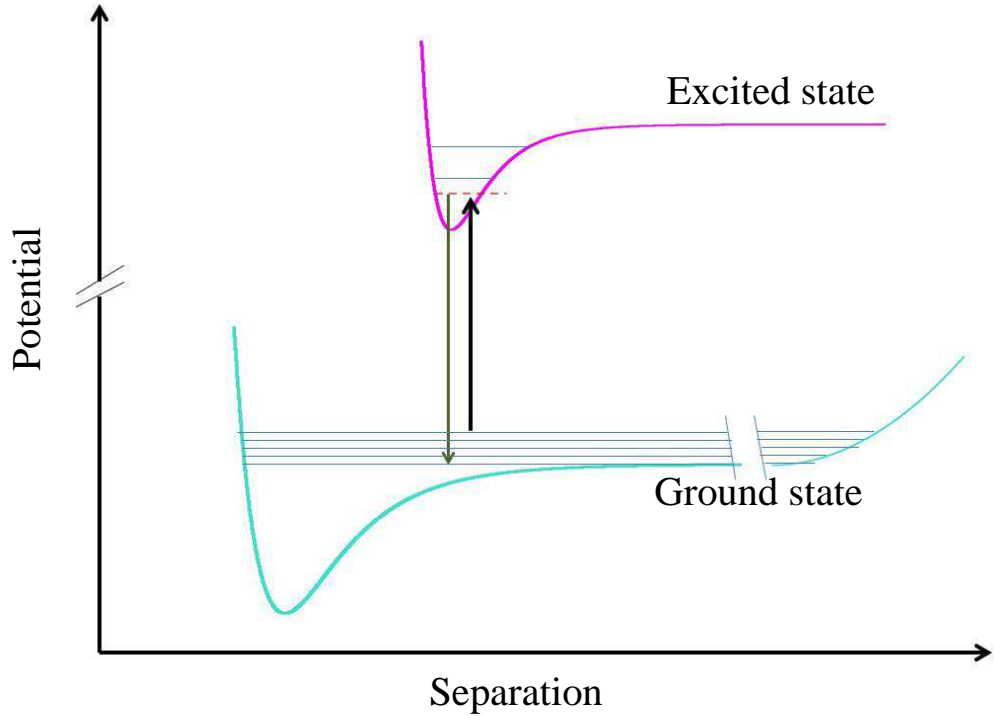}
 \caption{Schematic diagram showing molecular ground- and excited-state potentials for photoassociative
Raman sideband cooling. The horizontal axis represents the separation between the two atoms. Two
ground state atoms are assumed to be confined in a trap while the excited molecule may or may not
be trapped. Whether the excited molecule will remain trapped in the same trap that confines the two
ground-state atoms depends on the existence of a magic wave length for trapping both the atomic
and molecular states.}  
 \label{fig:fig1}
\end{figure}

However, if the two atoms are confined in a trap or optical lattice site or in a standing-wave node or antinode of a cavity field, then the motional states of the two atoms will be discrete and the difficulties associated with the continuum as pointed above may be removed. In case of singly trapped atoms or ions a well known sub-Doppler cooling technique is Raman sideband cooling. On the other hand, investigation and demonstration of single atom cavity cooling is also done. We here qualitatively discuss a two-photon cooling scheme of a pair of interacting atoms in a trap or cavity using photoassociative transitions between discrete motional states of 
the two ground-state atoms and the excited molecular states as schematically
depicted in Fig \ref{fig:fig1}. 

We have organized the paper in the following way. In section \ref{sect2}, a brief review in Raman sideband cooling is given. In section \ref{sect3}, a qualitative discussion is made on photoassociative Raman sideband cooling of a pair of atoms in a trap. In section \ref{sect4}, a brief review and qualitative discussion is made on photoassociative cooling of an atom pair in a cavity. The paper is concluded in section \ref{sect5}.

   \section{Raman sideband cooling: A brief review}
   \label{sect2}
Here we extend our idea of photoassociative cooling of Ref.\cite{pact} to two-photon sideband cooling of a tightly confined atom-pair in a trap.
In case of an atom or ion (which we generally call a species) in a 1D harmonic trap, its motional states can be characterized by the motional quantum number $n$ and energy $E_n=(n+1/2)\hbar \omega$ 
with trap frequency $\omega$. We consider two internal states of the system and denote by $|g,n_g\rangle$ and $|e,n_e\rangle$ where $|g(e)\rangle$ denotes internal ground (excited) states of the species. 
Let $\omega_0$ be the resonant transition
frequency between the two states and $\gamma$ be the spontaneous linewidth of the excited state. We assume that the species is tightly confined so that the length scale of the trap 
($\sqrt{\hbar/m\omega}$) is much smaller than the wavelength of the spontaneously emitted photon (LD regime). In that case if a laser beam is incident along the direction of atomic 
motion, the absorption spectrum shows a sideband structure which is composed of a central line at frequency $\omega_0$ and equi-spaced sideband lines on both side of it with a spacing 
of $\omega$. The sidebands can be resolved if $\omega\gg\gamma$. The intensity is maximum for frequency $\omega_0$ and is much larger than other sidebands. This indicates that a spontaneous emission 
from  $|e,n_e\rangle$ to $|g,n_g\rangle$ is the most favorable for $n_e=n_g$. This is a consequence of tight confinement of the species in the trap where the recoil energy of emission is much 
less than $\hbar\omega$ and upon emission it does not change its vibrational quantum number $n$. Thus, if we apply a laser which is tuned to resonance to the lower sideband 
($\omega_L=\omega_0-\omega$) then 
a stimulated absorption occurs from $|g,n\rangle$ to $|e,n-1\rangle$ which reduces one vibrational quantum of energy ($\hbar\omega$). It is then followed by spontaneous emission favorably 
to $|g,n-1\rangle$ completing a cooling cycle where energy is reduced by $\hbar\omega$. The method is known as resolved sideband cooling which is an efficient technique to cool atoms or ions beyond the Doppler cooling limit. 

The cooling limit can be shown to be $T_s=\hbar\omega/\left[k_B\ln\left(\omega^2/C_s\gamma^2\right)\right]$ \cite{wineland1987}, where $C_s$ is a constant that depends on the Clebsch-Gordon coefficient between the two internal states. So clearly smaller the $\gamma$ lower is the cooling limit. In Raman sideband cooling, a two photon Raman transition is used to cool a trapped species. The distinct feature of this technique is that in the formula for cooling limit as mentioned above, $\gamma$ is replaced by the linewidth of the stimulated Raman transition, which can be quite narrow. 

In our proposed method \cite{pact}, we have shown that CM motion can be trapped in optically generated potential wells. The depth of the wells depend on the total angular momenta of the two ground state atoms. So, for a pair of atoms having multiple total internal angular momenta, it is possible to generate different energy eigen states for CM motion for different internal states of the two atoms. Therefore, one can extend the technique of Raman sideband cooling of a single species to the case of CM motion of a pair of interacting atoms. Although it is difficult to trap more than one atom \cite{depue1999} by single-atom laser cooling technique due to light-assisted collision losses, nevertheless it is achievable \cite{winoto1999} with currently available technology of
cooling and trapping neutral atoms. Thus, the idea of sideband cooling can be extended further to a pair of atoms in a trap or in a doubly occupied lattice site of an optical lattice.

  \section{The model: Photoassociative Raman cooling}
  \label{sect3}
 Our model is schematically shown in Fig \ref{fig:fig1}. Two ground-state atoms confined in a harmonic trap are
being cooled by Raman sideband cooling using photoassociative transitions to a ro-vibrational level in the
excited molecular potential. The downward transition from the excited molecular level to a two-atom trapped
state may be spontaneous or stimulated. In case of stimulated transitions, we have coherent two-photon
photoassociative transitions in a $\Lambda$-type three-level system as discussed elsewhere [31]. The Hamiltonian for two identical atoms in a isotropic harmonic trap can be written as 

\begin{eqnarray}
 H(\vec{r}_1,\vec{r}_2)=\frac{1}{2m}\left(\vec{p}_1^2+\vec{p}_2^2\right)+\frac{1}{2} m\omega^2\left(\vec{r}_1^2+\vec{r}_2^2\right)+V\left(|\vec{r}_1-\vec{r}_2|\right)
\end{eqnarray}
where $V\left(|\vec{r}_1-\vec{r}_2|\right)=V(r)$ is the interatomic potential, $\omega$ is the harmonic frequency of the trap or the vibration frequency of a harmonically trapped atom. In terms of CM ($\vec{P}_{cm},\vec{R}_{cm}$) and relative ($\vec{p},\vec{r}$) coordinates it can be written as
\begin{eqnarray}
\label{isoho}
 H(\vec{R}_{cm},\vec{r})=\frac{\vec{P}_{cm}^2}{4m}+\frac{\vec{p}^2}{m}+ m\omega^2\vec{R}_{cm}^2+\frac{1}{4}m\omega^2\vec{r}^2+V\left(r\right)
\end{eqnarray}
Thus the CM and relative DOF are completely decoupled. What emerges from Eq ($\ref{isoho}$) is that there is a  CM harmonic trap of frequency $\sqrt{2}\omega$ and a relative trap of frequency $\omega/\sqrt{2}$ along with the short range atom-atom interaction potential $V(r)$.

For an anisotropic trap the system becomes more interesting because then CM and relative motion becomes coupled and energy exchange between these two DOF can occur. For small anisotropy the problem can be treated perturbatively in the isotropic basis. 
Interaction of light with such a system can be classified into two modes. One is the long range mode when the interaction between the atoms can be neglected and the
 excitation processes for the two atoms are almost independent. In the short range mode $V\left(r\right)$ plays an important role because in this case pair excitation 
can happen through PA. A similar sideband structure exists for such laser assisted dynamic process. On this basis, one can explore cooling of coupled DOF of the two atoms and the possibility of manipulating the interaction between them. 

\begin{figure}
 \includegraphics[width=0.8\textwidth]{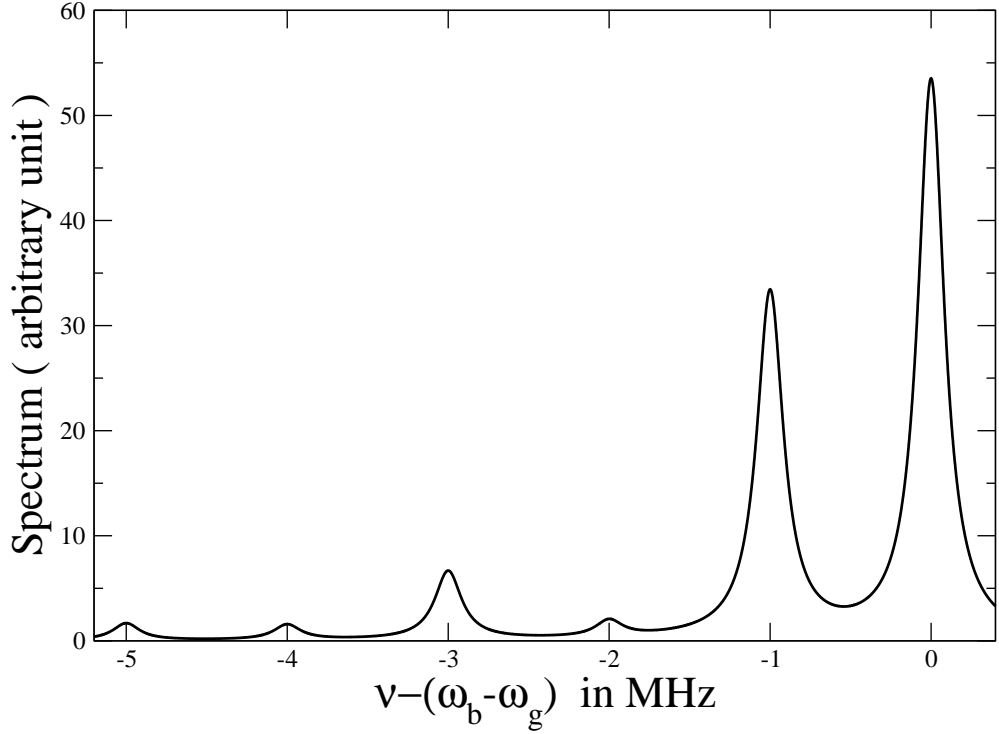}
 \caption{Emission spectra of the excited PLR state of $^{171}$Yb$_2$ molecule photoassociatively coupled to the two-atom trapped states. Along the horizontal axis is the frequency $\nu$ of the emitted photon. Here $\gamma=$364 KHz and $\omega=$1 MHz which approximately resolve the side bands. The PLR state chosen has the molecular vibration quantum number $v=$2. The height of each line is proportional to the square of the FC factor between the PLR state and a motional state ($n_r$) of the atom-pair in the trap.}  
 \label{sidebandPA}
\end{figure}

We consider a pair of fermionic $^{171}$Yb atoms which have $p$-wave PLR excited states. The pair of atoms is confined in a harmonic trap of frequency $\omega=1$ MHz for which
the trap size is $\sqrt{\hbar/\mu\omega}\sim 1 $ micrometer. For simplicity, we have considered an isotropic harmonic trap. We choose the excited PLR state with molecular vibration quantum number $v=$2. The fluorescent spectrum of an excited PLR state depends on the Franck-Condon (FC) overlap factor between the excited state and different ground vibrational states of the trap which are characterized by the radial quantum number $n_r$ of the isotropic harmonic trap in relative coordinate space. Since the spontaneous linewidth of the excited state is 364 KHz which is three times smaller than the trap frequency a resolved sideband structure is formed as shown in Fig \ref{sidebandPA}. If the atom-pair is excited from a higher motional state i.e. the frequency of the applied laser is tuned to lower sidebands then it will decay predominantly to the lowermost motional state emitting a photon of higher frequency as is clear from Fig \ref{sidebandPA}. This is because it has the maximum probability of spontaneous transition to the lowermost vibrational state. This indicates the possibility of cooling of relative motion. Alternatively, two-photon photoassociative Raman transition may be employed to cool the motional DOF.

   \section{Idea of photoassociative cavity cooling}
  \label{sect4}
We here give a brief review of some recent development in cavity cooling of atoms. We further qualitatively discuss the possibility of extending this technique to cool motional kinetic energy of a pair of interacting atoms confined in an antinode of a standing wave of cavity fields. As previously discussed we know that all conventional methods of laser cooling of atoms rely on the repeated optical pumping cycles followed by the spontaneous emission of the photons from the atom. Hence this type of cooling can be applied only to certain atomic species which shows closed optical transition. On the other hand the cavity cooling does not rely on spontaneous emission. In this case, a single atom is strongly coupled to a high-finesse cavity and the spontaneous emission of previous cooling schemes is replaced by emission of photon from the cavity. By cavity cooling we can cool systems that cannot be cooled via conventional cooling schemes, such as molecules as they do not posses the closed transition.

In 1997, Peter Horak and others \cite{horakprl1997,ritschpra1998} had presented a cavity cooling mechanism. They have shown that the strong atom-field coupling causes the resonance frequency of the cavity being shifted depending on the particle position. Hence, the single particle induces a position and time dependent index of refraction in the cavity. This cooling scheme is closely related to Doppler cooling, but the Sisyphus effect here arises due to the change of intracavity photon number correlated with atomic position, on the other hand in the standard Doppler cooling scheme the intensity of the cooling fields are fixed. The cavity decay rate $\kappa$ plays the same role here as of spontaneous emission in usual Doppler cooling. The 
achievable steady temperature for this case is, $k_B T = \hbar \kappa$. Hence for appropriately chosen parameters temperature much below the Doppler cooling limit can 
be accessed for long trapping times of the atoms, positioned at exactly anti-nodes of the field. Collective-emission-induced velocity-dependent light forces on atoms inside a resonator is experimentally
 observed by V Vuleti\'{c} {\it et al.}\cite{vuleticprl2003} in 2003. They cooled a falling sample of 3 $\times$ 10$^6$ cesium atoms to a temperature as low as 7 
$\mu$K, well below the free-space Doppler-limit. Rempe {\it et al.}\cite{rempenature2004} had shown cavity cooling of a single $^{85}$Rb atom stored 
in a intracavity dipole trap. In quantum information processing with trapped ions or atoms it is necessary to cool the motion of ions without disturbing the internal state where the information
 is stored. Cavity cooling allows to cool an atom without destroying the quantum superposition of the internal state. 
 
 There are two types of cavity cooling : cavity Doppler cooling and cavity sideband cooling. Cavity Doppler cooling 
is the cooling procedure discussed above, cavity sideband cooling is used to cool free particles when they are strongly confined in LD regime 
\cite{cavitysideband}. Chuah {\it et al.} \cite{chuahpra2013} investigate the dynamics of cavity cooling of a single ion $^{138}$Ba$^+$ beyond the LD regime. They
 demonstrate the cooling to sub Doppler temperature. Leibrandt {\it et al} has observed cavity cooling for single trapped $^{88}$Sr$^{+}$ ion in the 
resolved-sideband regime \cite{vuleticprl2009}. They also predict 
that this can be used to cool large molecular ions to motional ground state. 

Motivated by these recent developments in cavity sideband cooling of single atoms, we believe that
our photoassociative cooling idea can be extended to pairs of interacting atoms in a cavity. In a two-mode
cavity, one can treat both Stokes (photoassociation radiation) and the anti-Stokes fields quantum mechanically,
or in a single-mode cavity one can use the cavity mode as that of the anti- Stokes field while photoassociative
transition may be driven by an external laser. This will open a new perspective in cavity quantum optics - one
can explore new quantum optical phenomena at an interface between atomic and molecular states.  
    \section{conclusion}
  \label{sect5}
   In conclusion, we have qualitatively discussed how our recently proposed scheme of photoassociative
cooling and trapping (PACT) of interacting atoms can be extended to photoassociative Raman sideband
cooling of a pair of atoms trapped in Lamb-Dicke regime. The experimental realization of such cooling effect
will open a new perspective for many-body physics of interacting atoms in an optical lattice. Because, our
method can not only cool and trap the CM motion of two atoms, but it is capable of coherently manipulating interactions between atoms via optical Feshbach resonance. Further research and development of our
proposed method may lead to completely new route for simultaneously cooling, trapping and achieving
quantum degeneracy in an interacting Bose or Fermi atomic gas.

\vspace{1.0cm}

\end{document}